\documentclass[twocolumn,times,tighten]{aastex631}

\usepackage{xspace}
\usepackage[T1]{fontenc}

\newcommand{\FeKa}{Fe K\ensuremath{\alpha}\xspace}

\newcommand{\kms}{\ensuremath{\mathrm{km\ s^{-1}}}\xspace}
\newcommand{\NH}{\ensuremath{N_{\mathrm{H}}}\xspace}

\newcommand{\xmm}{{XMM-Newton}\xspace}

\newcommand{\swift}{{Swift}\xspace}

\newcommand{\hst}{{HST}\xspace}

\newcommand{\nustar}{{NuSTAR}\xspace}

\newcommand{\ergs}{{\ensuremath{\rm{erg\ s}^{-1}}}\xspace}

\newcommand{\cm}{{\ensuremath{\rm{cm}^{-2}}}\xspace}

\newcommand{\spex}{{\tt SPEX}\xspace}

\newcommand{\xil}{\ensuremath{{\log \xi}}\xspace}

\newcommand{\lya}{Ly\ensuremath{\alpha}\xspace}

\newcommand{\civ}{\ion{C}{4}\xspace}

\newcommand{\mrk}{{Mrk 841}\xspace}

\newcommand{\dbb}{{\tt dbb}\xspace}
\newcommand{\comt}{{\tt comt}\xspace}
\newcommand{\pow}{{\tt pow}\xspace}
\newcommand{\refl}{{\tt refl}\xspace}
\newcommand{\pion}{{\tt pion}\xspace}
\newcommand{\hot}{{\tt hot}\xspace}

\usepackage{xcolor}

\usepackage{xfrac}
\received{2023 April 27}
\accepted{2023 June 20 by ApJL}
\shorttitle{Dimming of Continuum Captured in Mrk 841: New Clues on the Nature of the Soft X-ray Excess}
\shortauthors{Mehdipour et al.}
\graphicspath{{./}{figures/}}
\begin{document}

\title{\large Dimming of Continuum Captured in Mrk 841: New Clues on the Nature of the Soft X-ray Excess}

\author[0000-0002-4992-4664]{Missagh Mehdipour}
\affiliation{Space Telescope Science Institute, 3700 San Martin Drive, Baltimore, MD 21218, USA; \href{mailto:mmehdipour@stsci.edu}{mmehdipour@stsci.edu}}

\author[0000-0002-2180-8266]{Gerard A. Kriss}
\affiliation{Space Telescope Science Institute, 3700 San Martin Drive, Baltimore, MD 21218, USA; \href{mailto:mmehdipour@stsci.edu}{mmehdipour@stsci.edu}}

\author[0000-0001-5540-2822]{Jelle S. Kaastra}
\affiliation{SRON Netherlands Institute for Space Research, Niels Bohrweg 4, 2333 CA Leiden, the Netherlands}
\affiliation{Leiden Observatory, Leiden University, PO Box 9513, 2300 RA Leiden, the Netherlands}

\author[0000-0001-8470-749X]{Elisa Costantini}
\affiliation{SRON Netherlands Institute for Space Research, Niels Bohrweg 4, 2333 CA Leiden, the Netherlands}
\affiliation{Anton Pannekoek Institute, University of Amsterdam, Postbus 94249, 1090 GE Amsterdam, The Netherlands}

\author[0000-0001-7557-9713]{Junjie Mao}
\affiliation{Department of Astronomy, Tsinghua University, Haidian DS 100084, Beijing, People’s Republic of China}
\affiliation{SRON Netherlands Institute for Space Research, Niels Bohrweg 4, 2333 CA Leiden, the Netherlands}

\begin{abstract}
We report on a remarkable change in the spectral energy distribution (SED) of Mrk 841, providing new insights on how the soft X-ray excess emission in active galactic nuclei (AGNs) is produced. By Swift monitoring of a sample of Seyfert-1 galaxies, we found an X-ray spectral hardening event in Mrk 841. We thereby triggered our XMM-Newton, NuSTAR, and HST observations in 2022 to study this event. Our previous investigations of such events in other AGNs had shown that they are caused by obscuring winds. However, the event in Mrk 841 has different spectral characteristics and origin. We find it is the soft X-ray excess component that has become dimmer. This is, importantly, accompanied by a similar decline in the optical/UV continuum, suggesting a connection to the soft X-ray excess. In contrast, there is relatively little change in the X-ray power-law and the reflection components. Our SED modeling suggests that the soft X-ray excess is the high-energy extension of the optical/UV disk emission, produced by warm Comptonization. We find the temperature of the disk dropped in 2022, explaining the observed SED dimming. We then examined the Swift data, taken over 15 years, to further decipher the UV and X-ray variabilities of Mrk 841. A significant relation between the variabilities of the X-ray spectral hardness and that of the UV continuum is found, again suggesting that the soft excess and the disk emission are interlinked. This is readily explicable if the soft excess is produced by warm Comptonization.
\end{abstract}
\section{Introduction} 
\label{sect_intro}

Outflows/winds in AGNs transport mass and energy away from the central engine and into the host galaxy. This may have important consequences for the co-evolution of supermassive black holes (SMBHs) and their host galaxies through the resulting feedback mechanism between the AGN activity and star formation (see e.g. \citealt{King15,Gasp20}). Thus, ascertaining the physical properties and energetics of AGN winds, and how they are launched and driven, are important for determining their role in AGNs and assessing their impact on their environment.

As outflows in AGNs are ultimately powered and driven by the energy released from the accretion process, their physics is interlinked to that of accretion and its associated continuum radiation, called the spectral energy distribution (SED). Outflows are photoionized by this intense ionizing SED, thus imprinting their spectral signatures on the UV/X-ray continuum as absorption and emission features. Therefore, for the study of ionized outflows, the SED needs to be determined. On the other hand, to derive the intrinsic SED, the underlying emission prior to any continuum absorption needs to be established. Thus, spectral modelings of the ionized outflows and the SED are interwoven. The SED model is required for both photoionization modeling of the outflows, and also disentangling the absorption components from the continuum components. Understanding changes in the SED is needed for both proper modeling of the winds and probing the accretion-powered emission in AGNs.

It is not fully understood how the observed SED in AGNs is produced. The nature and origin of the components of the SED, and their physical relation with each other, are uncertain. Convolution of these components across wide energy bands, and their alteration due to absorption, are complicating factors. In particular, an outstanding problem is the origin of a puzzling ``soft X-ray excess'' emission. The observed X-ray spectra of AGNs often display an excess emission component in addition to the underlying X-ray power-law continuum at energies ${< 2}$ keV. There are different explanations proposed in the literature for the presence of the soft X-ray excess: relativistically-blurred ionized disk reflection (e.g. \citealt{Ros05,Cru06}); Compton up-scattering of the disk photons in a warm and optically-thick corona (e.g. \citealt{Done12,Kubo18}); non-thermal particle acceleration processes, including shocks or magnetic reconnection (e.g. \citealt{Fuku16}); consequences of relativistcally-blurred absorption by disk winds (e.g. \citealt{Gie04}). Determining the nature and origin of the soft X-ray excess is important as different interpretations ultimately have important implications for our understanding of the accretion and wind phenomena in AGNs.

Spectral variability is a key characteristic of AGNs, which is useful for probing the unknown properties of both the accretion/SED and the outflows. In changing-look AGNs, where there is a transformation of the intrinsic continuum, it allows us to probe major changes in the accretion activity and the associated radiation \citep{LaMa15,LaMa17,Mehd22a}. Likewise, in AGNs that undergo transient obscuration, this variability allows us to study different types of winds, and how they interact and influence each other \citep{Arav15,Kris19b,Dehg19b,Laha21}. 

Following the discovery of a connection between transient X-ray obscuration and broad-line region (BLR) winds in NGC~5548 \citep{Kaas14}, the \swift observatory has been utilized to monitor the X-ray spectral hardness variability in a sample of Seyfert-1 AGNs, in order to trigger joint target-of-opportunity (ToO) observations with \xmm, \nustar, and the Hubble Space Telescope's (HST) Cosmic Origins Spectrograph (COS). As a result, additional transient obscuring winds have been found in other AGNs, notably NGC~3783 \citep{Mehd17,Kris19}, Mrk~335 \citep{Long19,Park19} NGC~3227 \citep{Mehd21,Wang22,Mao22a}, and~MR 2251-178 \citep{Mao22}. 

In late December 2021 a new spectral hardening event was discovered in \mrk, which led to follow-up observations with \xmm, \nustar, and \hst in January and February 2022. However, unlike our previously triggered objects, the event in \mrk turned out to be very different. \mrk is a Seyfert-1 galaxy at redshift ${z = 0.03642}$ \citep{Falc99}. It was one of the first AGNs where the presence of the soft X-ray excess was discovered \citep{Arna85,Petr07}. 
The soft X-ray excess in \mrk was studied by \cite{Petr07} using \xmm/EPIC-pn data taken in 2001 and 2005. They tested different models for the soft excess, namely the relativistically-blurred reflection and absorption models. They found both models are able to reproduce the soft excess with statistically equivalent results. Such studies highlight the difficulty in the interpretation of the soft excess when considering only the X-ray band. However, inclusion of optical/UV data, and importantly their variability, are highly useful in identifying the right explanation for the soft excess. As we present in this paper, our modeling of the new broadband SED captured by the 2022 event, and the comparison with the historical SED, shines new light on the origin and nature of the soft X-ray excess.

\section{Comparison of new and historical observations} 
\label{sect_data}
%

%
\begin{figure}
\centering
\resizebox{\hsize}{!}{\includegraphics[angle=0]{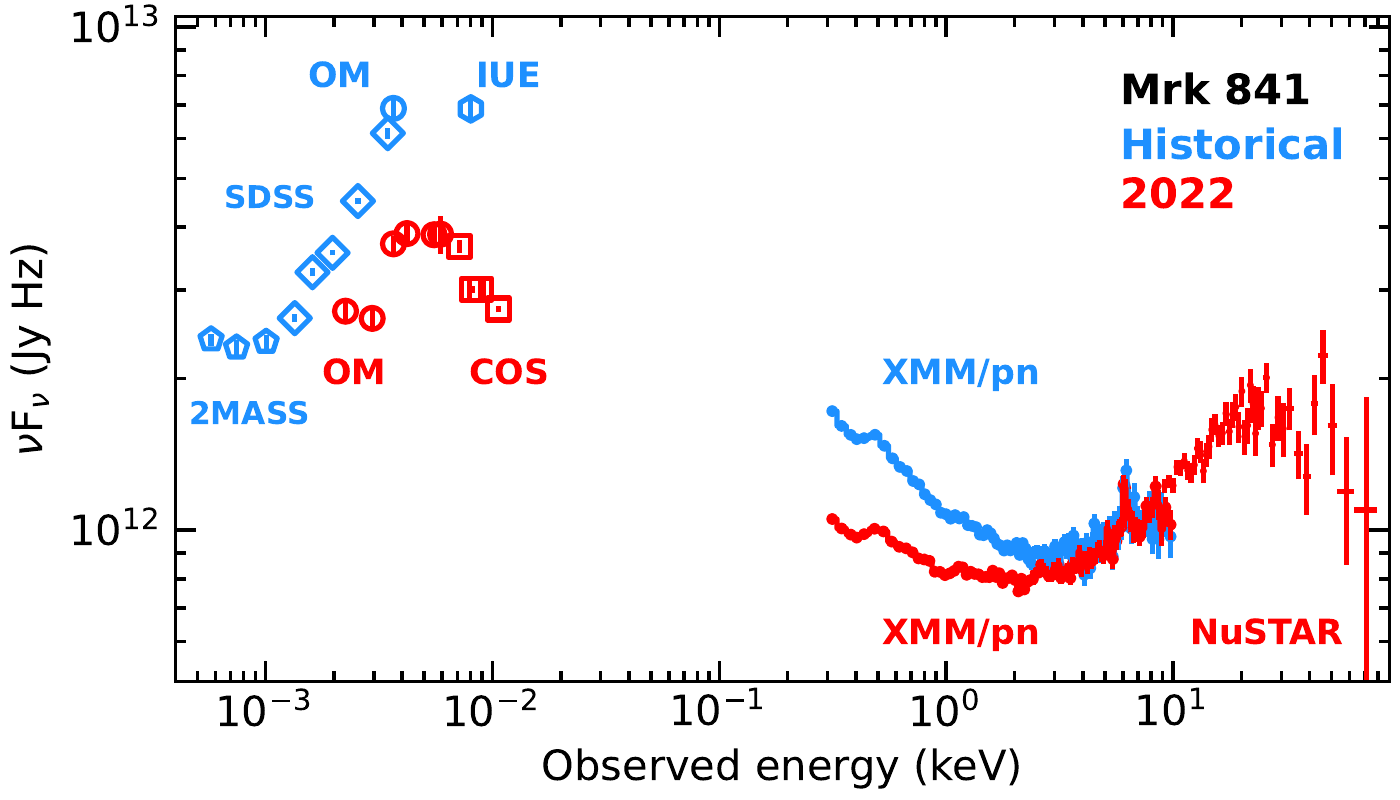}}
\caption{IR, optical, UV, and X-ray spectra of \mrk from the 2001/historical and 2022 epochs. The data are shown as observed, including all foreground effects. Major changes in shape and flux of the observed SED in 2022 are evident. The underlying intrinsic SEDs are derived and shown in Fig. \ref{fig_sed}.
\label{fig_spec}}
\vspace{0.1cm}
\end{figure}

Our \xmm and \nustar ToO observations of \mrk, as well as the \swift monitoring, were all approved as a \nustar Cycle 7 proposal. 
During the \swift monitoring of \mrk in December 2021, we first noticed an X-ray spectral hardening on 2021-12-26. Our triggered \nustar (50 ks) and \xmm (60 ks) ToO observations were then taken on 2022-01-09. 
Through a Director Discretion (DD) request, a HST/COS observation (1 orbit) was also taken on 2022-02-01 with the G130M and G160M gratings.

The \swift/XRT data products were obtained from the UK Swift Science Data Centre \citep{Evan07,Eva09}.
The \swift/UVOT and \nustar data were processed with HEASoft v6.30.1.
The \xmm data were reduced with SAS v20.0.0.
The HST/COS data of \mrk can be accessed via \dataset[10.17909/ajvk-5644]{\doi{10.17909/ajvk-5644}}.
The \xmm, \nustar, and HST/COS data are shown in SEDs of Figure \ref{fig_spec}.
The COS fluxes in this figure are from four bands that are free of spectral features to represent the far-UV (FUV) continuum.
In order to assess the SED changes that are evident in Figure \ref{fig_spec}, we compare the 2022 data with the historical observations.  
We describe below the data that were used for the construction of this historical SED of \mrk.

For our investigation we model the 2001 \xmm observation which was originally used in our simulations for the proposal to calculate the reference X-ray hardness ratio at the usual bright state of \mrk.
The 2001 \xmm/OM exposure was taken in the $U$ band only, whereas our 2022 OM data were taken in all six filters.
Thus, to help us with constraining the historical SED, we make use of other archival data published in the literature.
These data are not contemporaneous. However, in wavelength regions that are close or overlapping, the flux levels are similar. This suggests there is no substantial variability between these historical data. Nonetheless, the inclusion of these non-contemporaneous data is useful for the construction of the broadband SED.
We model SDSS data \citep{Adel08}, taken with the $u$, $g$, $r$, $i$, $z$ filters, to cover the optical and IR bands.
To check the SED at lower IR energies, we obtained 2MASS photometric data \citep{Skru06} taken with the $J$, $H$, $K_s$ filters.
To extend the historical SED as much as possible into the far-UV band, we also make use of an IUE flux measurement \citep{Bask05}, which like the 2001 \xmm observation is at the bright flux level of \mrk.
These archival data and the 2001 \xmm data, collectively referred to as ``historical'', are shown in Fig. \ref{fig_spec}.

\section{Spectral analysis and modeling} 
\label{sect_model}

We carried out our spectral modeling of \mrk in \spex v3.07.01 \citep{Kaa96,Kaas20}.
We fitted the 2022 and historical spectra jointly, allowing the possibility of coupling parameters for the two epochs.
All IR/optical, UV, and X-ray data shown in Figure \ref{fig_spec} were modeled, except the 2MASS data as they contain emission from the dusty torus (hence the rise in luminosity), which we are not studying here.
In our modeling, the redshift of \mrk was set to ${z = 0.03642}$ \citep{Falc99}, and the Milky Way absorption and reddening in our line of sight are taken into account using the \hot and {\tt ebv} models. 
The total Galactic \NH was fixed to $2.46 \times 10^{20}$~\cm \citep{Will13}, and the extinction to ${E(B - V) = 0.026}$ \citep{Schl11}.   
We use the extinction curve of \citet{Car89}, including the \citet{ODo94} update, with the ${R_V}$ ratio fixed to 3.1. 
The examination of HST and OM images shows that this object is a very compact source, and the starlight contamination by the host galaxy is minimal (few \%).

To derive the intrinsic broadband continuum, we apply models that have been previously tried and tested in modeling SEDs of other similar Seyfert-1 galaxies (see e.g. \citealt{Meh15a,Mehd21}).
We use a disk blackbody component (\dbb) to model the optical/UV continuum emission from the accretion disk.
To model the extreme UV (EUV) emission and the soft X-ray excess component, we use a Comptonization component (\comt).
In our modeling the \dbb component represents the outer region of the disk (without a warm-Comptonizing corona) and the \comt component represents the inner region of the disk (with a warm-Comptonizing corona). We thus couple the seed photon temperature of \comt to the maximum temperature of \dbb in our fits.
We use a power-law model (\pow) to model the underlying X-ray continuum and a neutral, narrow, reflection component (\refl) to fit the \FeKa line and the Compton hump at hard X-rays.
The parameters of the illuminating power-law for {\tt refl} are coupled to those of the 2001 power-law.
The high-energy exponential cut-off of the power-law was fixed to 300 keV \citep{Petr07}, and the low energy one to 1 Ryd.

\begin{figure}
\vspace{0.2cm}
\centering
\resizebox{\hsize}{!}{\includegraphics[angle=0]{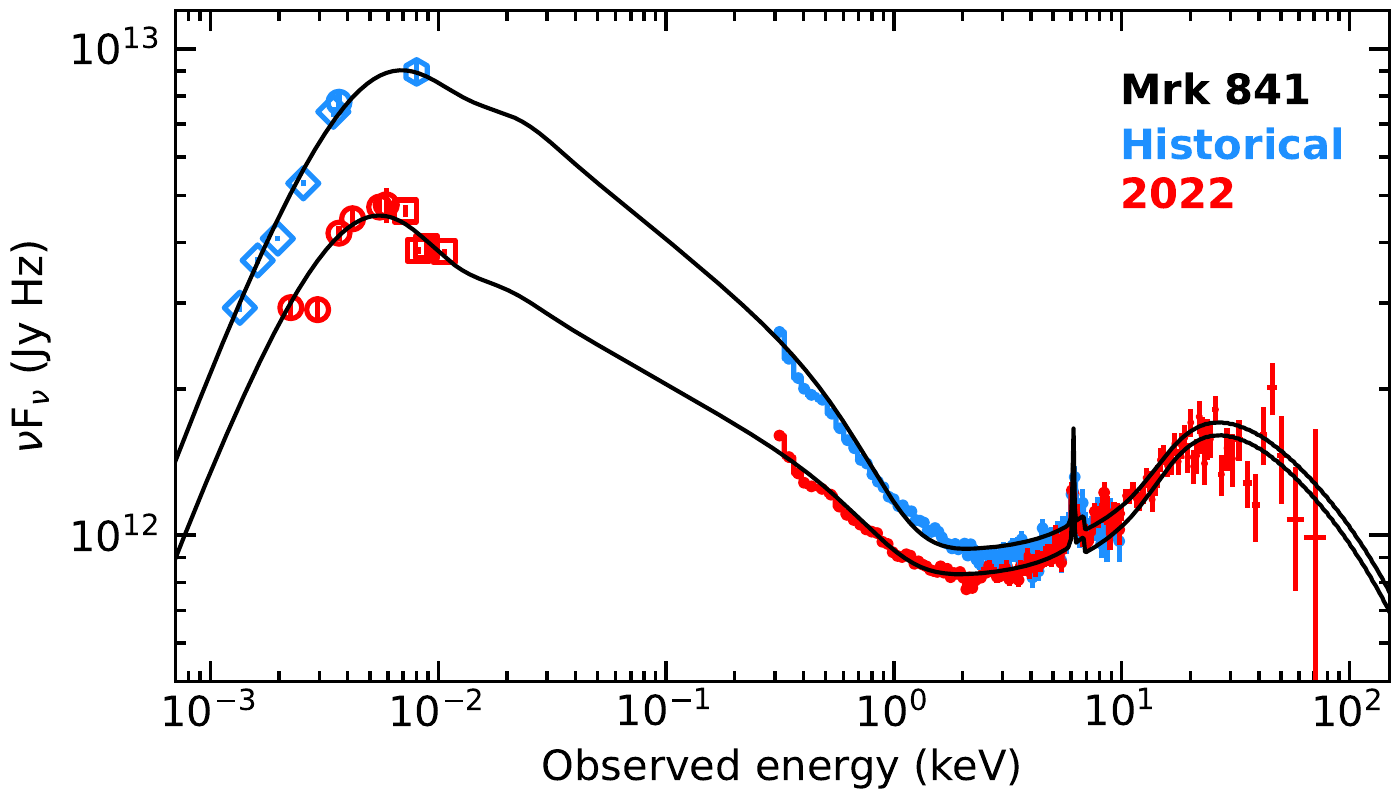}}
\resizebox{\hsize}{!}{\includegraphics[angle=0]{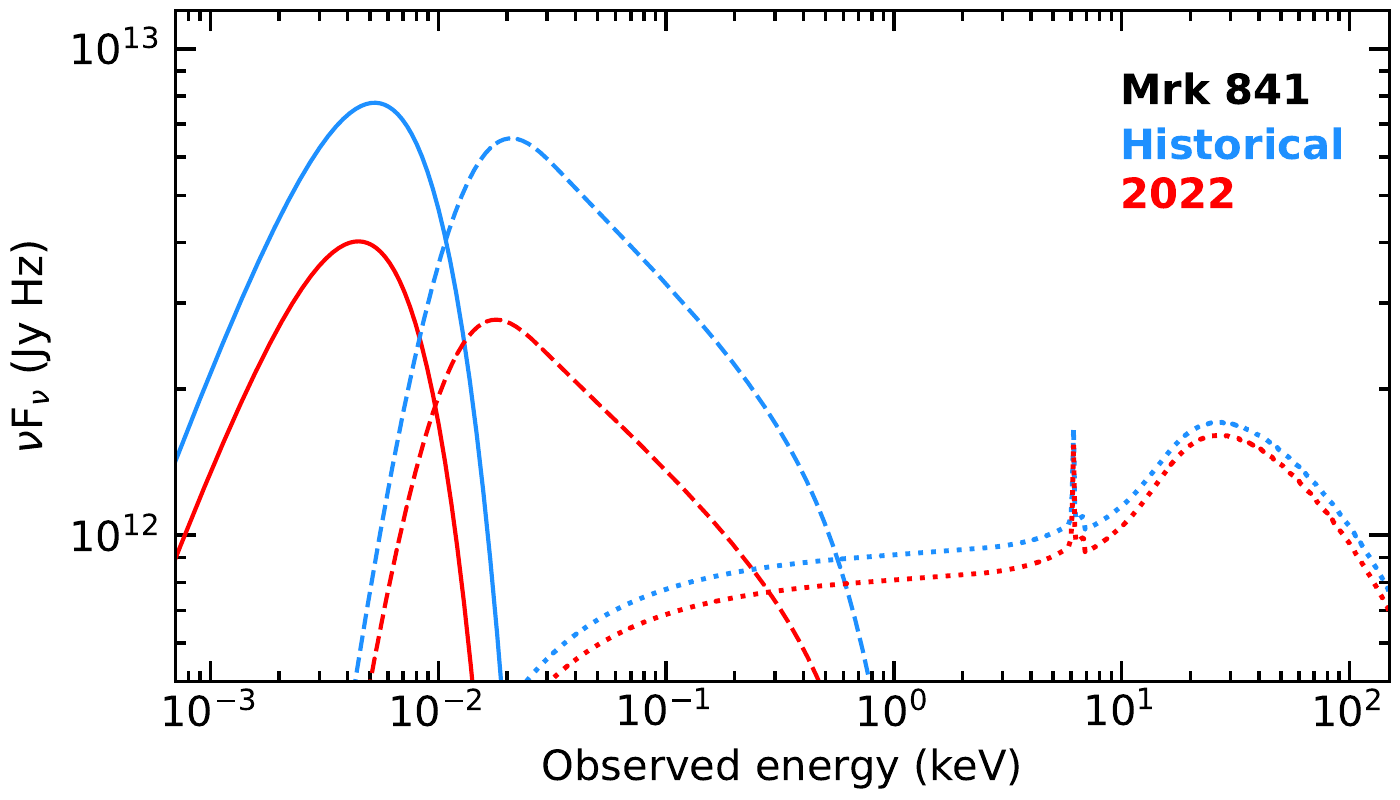}}
\vspace{-0.6cm}
\caption{Top panel: derived intrinsic SED continuum model for the 2001/historical and the 2022 epochs of \mrk. Bottom panel: individual components of the SED. The significant changes in the {\tt dbb} (solid line) and {\tt comt} (dashed line) components are apparent. The parameters of the continuum components are given in Table \ref{table_para}.
\label{fig_sed}}
\vspace{0.0cm}
\end{figure}

In our SED modeling we took into account the X-ray absorption by the warm absorber in \mrk, using the \pion model \citep{Meh16b}.
Our modeling of the \xmm/RGS data shows that the intrinsic absorption is rather weak, due to relatively low \NH and high ionization parameter $\xi$.
We find the warm absorber consists of two ionization components.
The first component has parameters of $\xil = 2.4$, ${\NH = 4.5 \times 10^{20}}$ \cm, with an outflow velocity of $-750$~\kms. The parameters of the second, more ionized, component are $\xil = 3.2$, ${\NH = 1.5 \times 10^{21}}$ \cm, outflowing at $-530$~\kms.
We find the parameters of the warm absorber are consistent with no change between the two epochs, except the ionization parameters which are lowered by 0.2 dex in response to the dimming of the ionizing SED in 2022.
Figure \ref{fig_sed} (top panel) shows our derived intrinsic SED models for the historical and the 2022 epochs. 
The individual components of the SEDs are shown in the bottom panel. 
The best-fit parameters of the continuum components are provided in Table \ref{table_para}.
We discuss the results and the SED changes in the following section.

%

%
\begin{deluxetable}{c | c c}
\tablecaption{Best-fit parameters of the continuum components of the SED, derived from modeling the historical and 2022 data of \mrk.
\label{table_para}}
\tablewidth{0pt}
\setlength{\tabcolsep}{10pt}
\tablehead{
Parameter & \colhead{Historical} & \colhead{2022}
}
\startdata
\multicolumn{3}{c}{Disk component ({\tt dbb}):} 						\\
Area	($10^{29}$ cm$^{2}$)   &  ${2.7 \pm 0.2}$ & ${2.7 \pm 0.2}$ \\
$T_{\rm max}$ (eV)           &	${4.5 \pm 0.1}$ & ${3.8 \pm 0.1}$ \\
\hline                     
\multicolumn{3}{c}{Soft X-ray excess component ({\tt comt}):} 						\\
Normalization	       &  ${8.1 \pm 0.1}$ & ${3.9 \pm 0.1}$ \\
$T_{\rm seed}$ (eV)  &	${4.5}$ (f) & ${3.8}$ (f) \\
$T_{\rm e}$ (keV)    &  ${0.18 \pm 0.02}$ & ${0.18 \pm 0.02}$ \\
Optical depth $\tau$ &	${19 \pm 3}$  & ${19 \pm 3}$ \\
\hline
\multicolumn{3}{c}{X-ray power-law component ({\tt pow}):} 						\\
Normalization			& ${1.8 \pm 0.1}$ & ${1.6 \pm 0.1}$ \\
Photon index $\Gamma$		& ${1.98 \pm 0.02}$	& ${1.98 \pm 0.02}$	\\
\hline
\multicolumn{3}{c}{X-ray reflection component ({\tt refl}):} 						\\
Reflection fraction			& ${0.9 \pm 0.1}$ & ${0.9 \pm 0.1}$					\\
\hline
C-stat\,/ d.o.f.			& 1096\,/\,765 & 1241\,/\,918					\\
\enddata
\tablecomments{The normalization of the Comptonization component ({\tt comt}) is in units of $10^{55}$ photons~s$^{-1}$~keV$^{-1}$. The power-law normalizations of the {\tt pow} and {\tt refl} components are in units of $10^{52}$ photons~s$^{-1}$~keV$^{-1}$ at 1 keV.}
\vspace{-1cm}
\end{deluxetable}

\vspace{-0.3cm}
\section{Discussion and conclusions} 
\label{sect_discus}

Understanding the formation of continuum emission, across the electromagnetic spectrum, is important for both the studies of accretion (disk) and ejection (wind).
The unobservable EUV band is the least known region of the SED, and the soft X-ray excess is the most uncertain component.
Yet, the EUV band is arguably the most significant part of the SED as the luminosity is generally dominant in this band.
Ascertaining the bolometric luminosity of an AGN is dependent on the modeling of the uncertain EUV band. 
Importantly, the ionization and thermal structure of AGN winds, derived from photoionization modeling, is highly dependent on the ionizing EUV and the soft X-ray continuum. 
Thus, establishing global models that connect the optical/UV to the soft X-rays in a physically consistent fashion is crucial for both correct modeling of winds, as well as for understanding the physical association between the disk and the X-ray emitting coronas in AGNs.
The warm Comptonization interpretation of the soft X-ray excess \citep{Done12,Petr18,Kubo18} can provide a viable model for the EUV continuum.

The change in the SED of \mrk, captured with our ToO observations in 2022, provides useful clues into the nature of the soft X-ray excess.
The comparison of the new and historical spectra (Fig. \ref{fig_spec}) reveals a clear decline in both the soft X-ray and optical/UV bands, while the \FeKa line and the hard X-rays show little change.
So the soft X-ray excess and the optical/UV continuum became fainter together in 2022.
This is evident in both the raw data (Fig. \ref{fig_spec}) and the derived continuum models (Fig. \ref{fig_sed}).
For the 2001/historical SED, the intrinsic luminosity of the disk blackbody component ({\tt dbb}) was ${4.9 \times 10^{44}}$ \ergs and that of the Comptonization component ({\tt comt}) ${5.3 \times 10^{44}}$ \ergs.
This slightly higher luminosity of \comt compared to \dbb merely means that the inner part of the disk (which has a warm-Comptonizing corona) is more luminous than the outer part of the disk (which does not have a warm-Comptonizing corona).
Our SED modeling shows that the intrinsic luminosities of both the {\tt dbb} and {\tt comt} reduced by half in 2022.

In the 2001 \xmm observation, the intrinsic luminosity of the X-ray power-law over 0.3--10 keV was ${1.0 \times 10^{44}}$ \ergs, while in the 2022 observation this luminosity was smaller by only 10\%. 
Even if the entire luminosity of the X-ray corona was absorbed and reprocessed to make the soft X-ray excess, it would be energetically insufficient by a factor of five.
Furthermore, the luminosities of the reflection component (\FeKa line) are consistent with no change.
While hard X-ray \nustar data are not available for the historical observation, the \FeKa line and the continuum up to 10 keV still provide useful measurements for comparison.  
The luminosity change in the soft excess component is larger by more than an order of magnitude than that of the power-law, also making it inconsistent for the power-law to power the luminosity change of the soft excess in \mrk.
Warm Comptonization is the only energetically plausible model that can explain the joint dimming of the optical/UV and the soft excess, as according to this model the soft excess is the extension of the optical/UV disk component into the EUV and soft X-ray bands.
This reasoning for the origin of the soft excess goes beyond considerations of the statistical quality of X-ray-only fits as different models can provide similarly good fits to the soft excess.

In the X-ray illuminated disk models, the X-ray power-law emission from a hot corona is reprocessed by other regions, such as the disk or the torus.
These X-ray models provide predictions for the spectral properties and timing/lag characteristics of the continuum (power-law) and the reflected components (soft excess, \FeKa line, and the Compton hump); see e.g. \citet{Ros05,Garc19}.
However, from the X-rays alone, it is challenging to distinguish between the warm Comptonization and the reflection models for the soft excess (e.g. \citealt{Garc19}). 
This is because with either model, a combination of parameters can produce equally good fits to the X-ray spectra, albeit sometimes with extreme conditions in both scenario, such as a maximally spinning black hole to blur out the reflected emission features from the disk, or applying a non-standard corona for the warm Comptonization scenario. 
Understanding how the disk and different X-ray coronas may interact and impact each other (e.g. \citealt{Ball20,Dovc22}) are important for better understanding the nature of the soft excess.
Importantly, the UV continuum's luminosity, shape, and variability characteristics can provide additional valuable information, which sometimes are not taken into account by some X-ray studies in the literature.
In the case of \mrk, the reflection explanation for the soft excess is unlikely to explain the large, simultaneous dimming of the soft excess and the optical/UV continuum.
The luminosity of the X-ray power-law ({\tt pow}) changed little (10\%), and that of the X-ray reflection component ({\tt refl}) is consistent with no change.
The \FeKa line and the Compton hump are fitted well with the same neutral, narrow, {\tt refl} component, suggesting that it does not follow the behavior of the soft excess. 
We note that warm Comptonization and blurred reflection are not mutually exclusive, and both can occur and contribute to the soft excess of an AGN, but to what extent is the question.
The \mrk results suggest that at least the bulk of the soft excess emission, and its variability, are most plausible with the warm Comptonization scenario \citep{Nand21,Kawa23,Mitc23}.

\begin{figure}
\vspace{0.1cm}
\centering
\resizebox{0.97\hsize}{!}{\includegraphics[angle=0]{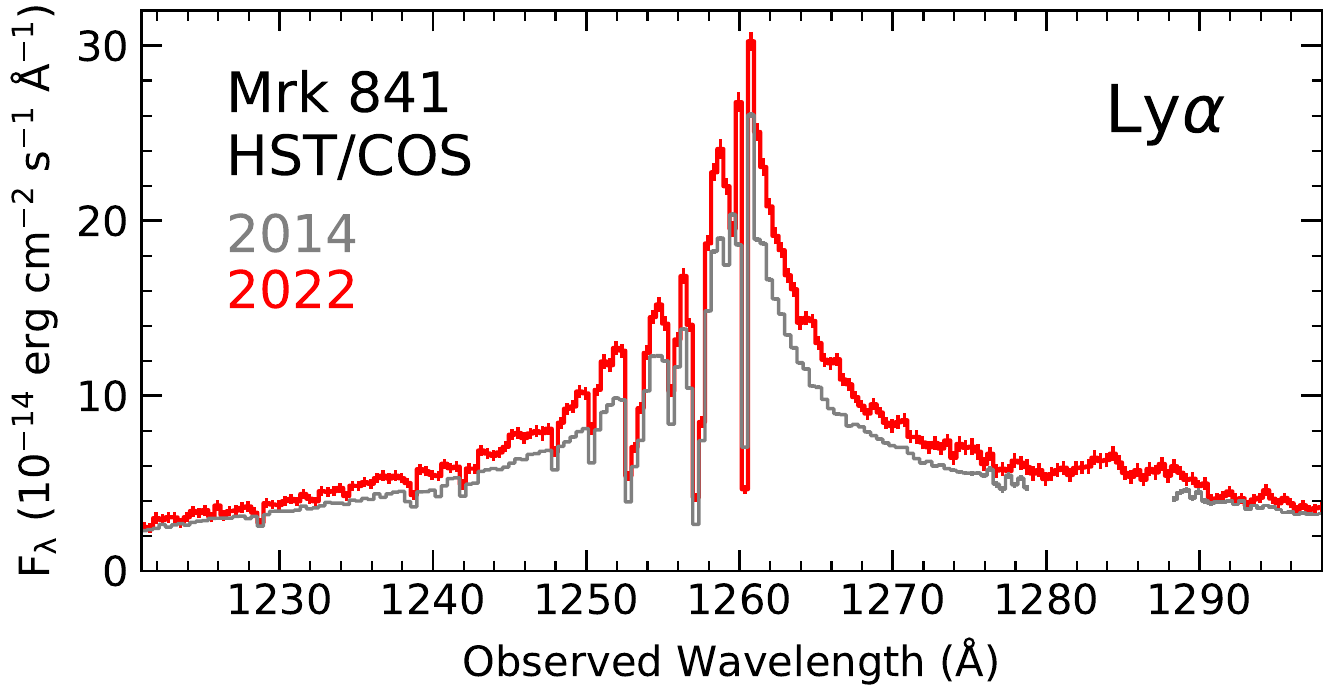}}
\resizebox{0.97\hsize}{!}{\includegraphics[angle=0]{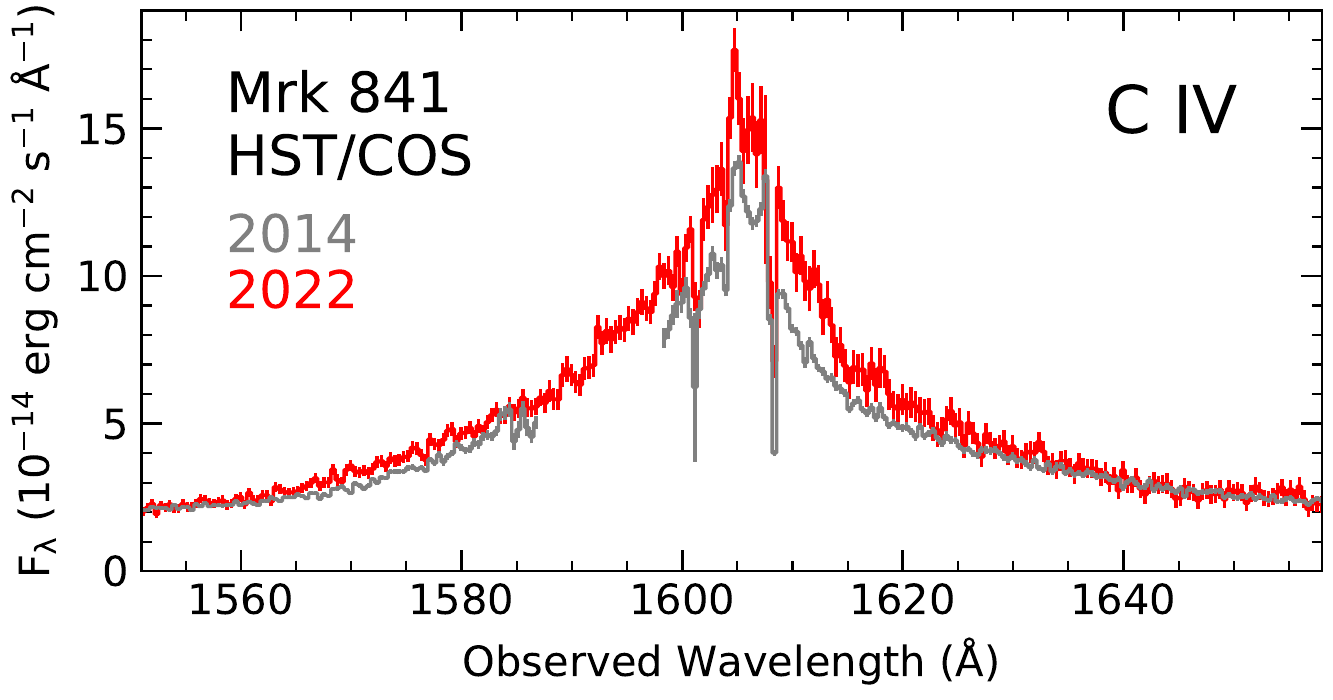}}
\vspace{-0.2cm}
\caption{Comparison of the new (2022) and 2014 HST/COS spectra of \mrk in the \lya (top panel) and \civ (bottom panel) regions. No broad, blueshifted, UV absorption features appear in the new data, consistent with our no-obscuration interpretation. The gaps in the 2014 spectrum are due to the instrumental setup of that observation.
\label{fig_hst}}
\vspace{-0.1cm}
\end{figure}

The studies of Seyfert-1 AGNs that underwent transient X-ray obscuration show that the optical/UV continuum does not become dimmer during these events.
For example, this is evident in the case of NGC~5548 \citep{Kaas14,Mehd22c} and NGC~3783 \citep{Mehd17}.
However, unlike these obscured AGNs, in \mrk the optical/UV flux declined significantly during the event (Figure \ref{fig_spec}).
Importantly, in those obscured AGNs, broad, blueshifted, UV absorption lines (namely \lya and \civ) appeared during the obscuration events. 
Figure \ref{fig_hst} shows a comparison of the 2022 and 2014 HST/COS spectra at the \lya and \civ regions.
It is clear that no broad absorption lines have appeared in 2022, and the emission line profiles are similar in 2014 and 2022. 
Finally, in addition to the above reasons, by fitting an X-ray obscuration component a significantly worse fit is found, and more importantly, such a component cannot explain the drop in the optical/UV band.
Therefore, it is conclusive that the 2022 event in \mrk is not a transient obscuration event.

%

%
\begin{figure}
\centering
\vspace{-0.2cm}
\resizebox{\hsize}{!}{\includegraphics[angle=0]{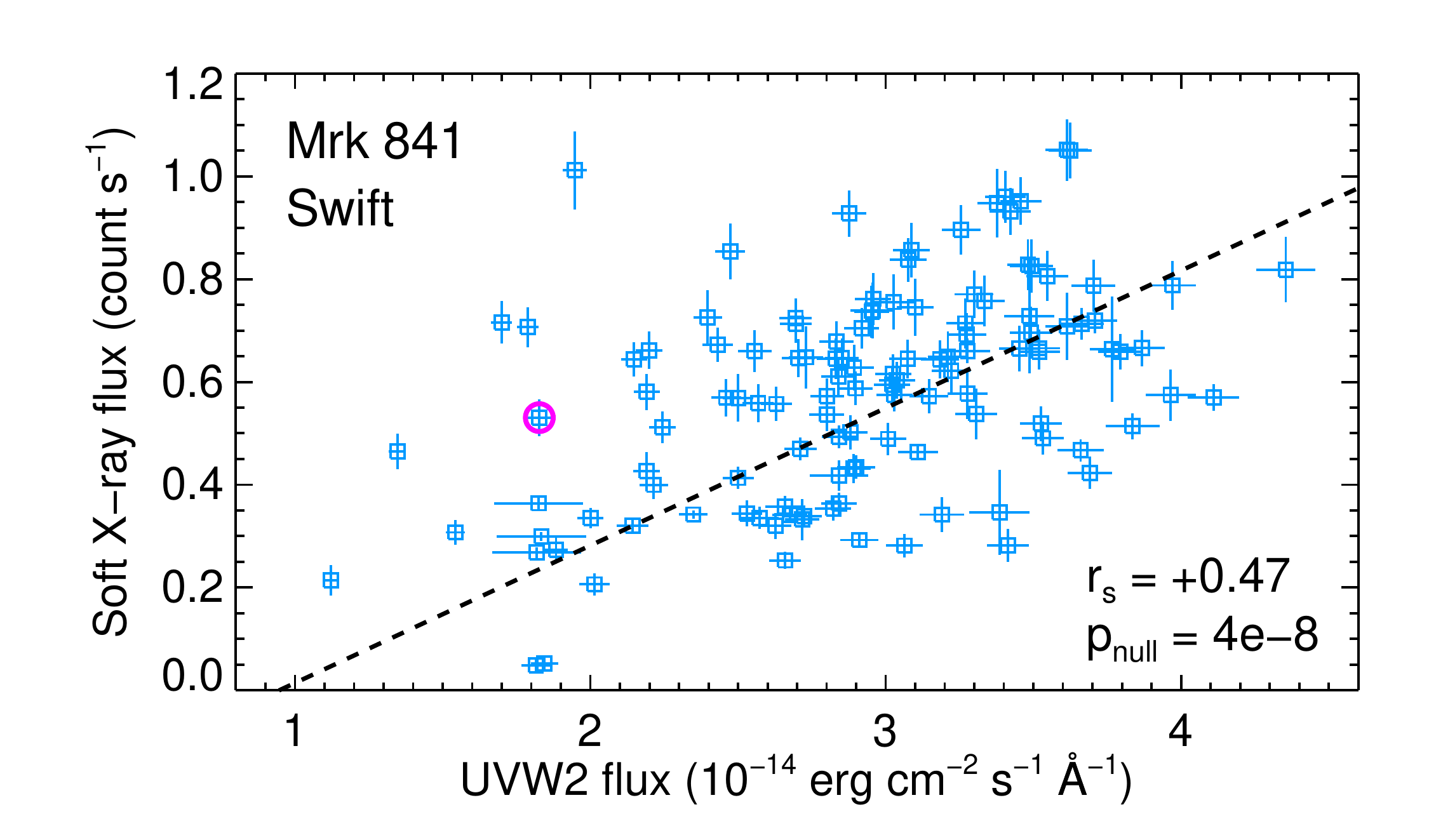}}\vspace{-0.5cm}
\resizebox{\hsize}{!}{\includegraphics[angle=0]{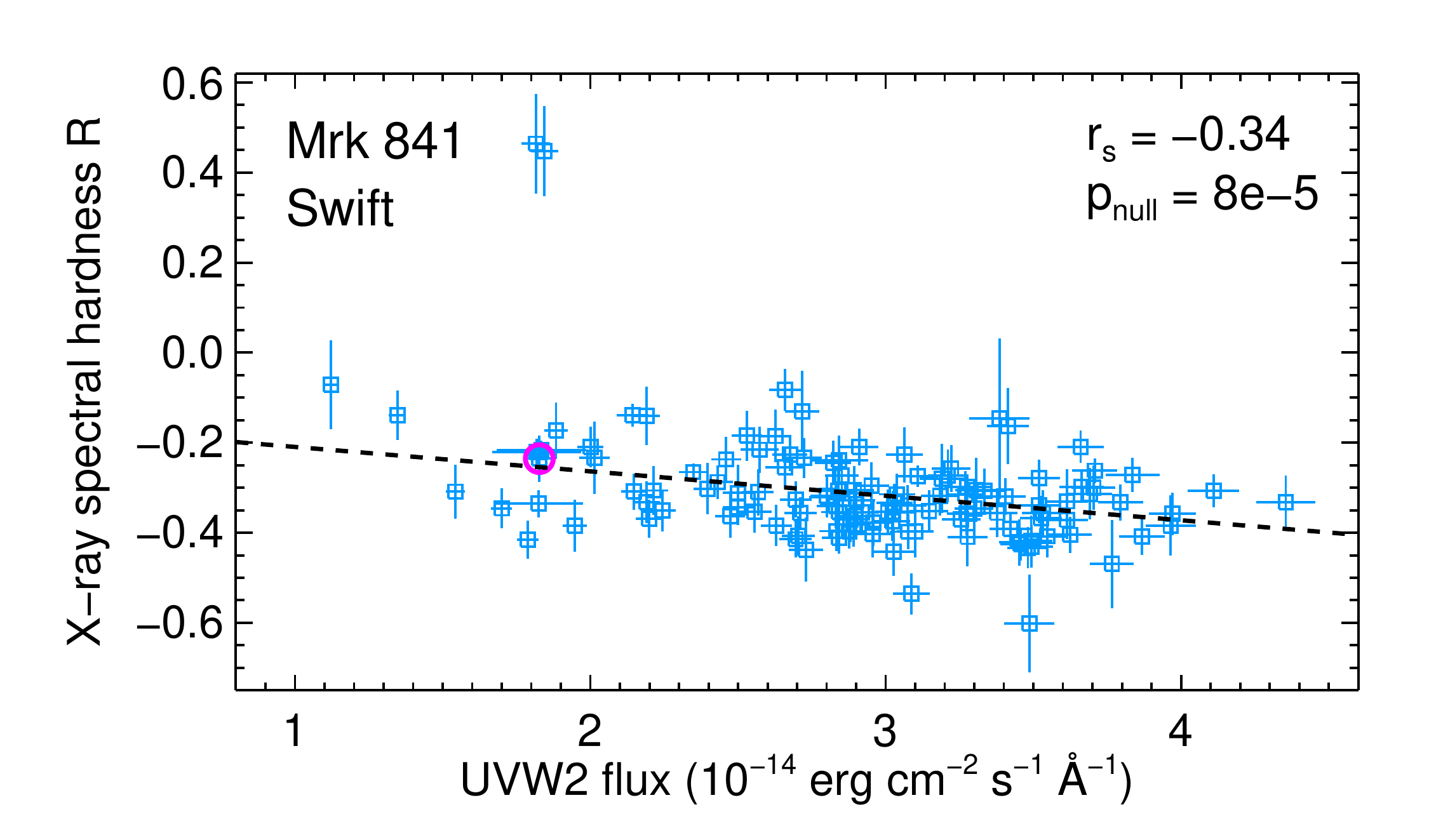}}\vspace{-0.2cm}
\caption{\swift/XRT soft X-ray flux (top panel) and the X-ray hardness ratio $R$ (bottom panel), plotted versus the \swift/UVOT UVW2 filter flux. The X-ray spectral hardness ratio ($R$) is defined as ${R = (H - S )/(H + S)}$, where $H$ and $S$ are the Swift XRT count rate fluxes in the hard (1.5--10 keV) and soft (0.3--1.5 keV) bands, respectively. The displayed data span from 2007 to 2022. The data corresponding to the January 2022 ToO event are indicated with a magenta circle. The Spearman’s rank correlation coefficient ($r_s$) and the corresponding null hypothesis p-value probability ($p_{\rm null}$) from the two-sided t-test are given in each panel. There is a significant positive correlation between the soft X-ray and UV fluxes (top panel), and a significant negative correlation between the hardness $R$ and the UV flux (bottom panel). 
\label{fig_swift}}
\vspace{0.0cm}
\end{figure}

\swift monitoring provides a useful probe of the X-ray and UV variabilities in \mrk.
Figure \ref{fig_swift} shows the soft X-ray flux plotted versus the UV flux (top panel), and the X-ray spectral hardness ratio $R$ versus the UV flux (bottom panel).
The \swift data span from 2007 to 2022.
The data show that there is a highly significant positive correlation between the soft X-ray and UV fluxes (top panel).
More interestingly, there is a highly significant negative correlation between the spectral hardness $R$ and the UV flux (bottom panel). 
This means as the source becomes brighter in UV, the X-ray spectrum becomes softer, i.e. the soft excess becomes brighter.
This important \swift finding independently verifies the results of our more detailed SED modeling (Figure \ref{fig_sed}). 
It demonstrates that the soft excess is directly connected to the UV continuum, consistent with the warm Comptonization explanation of the soft excess.
Our soft excess finding in \mrk is similar to that in Mrk 509 \citep{Meh11}. In that AGN, there was also a significant relationship between the variabilities of the optical/UV continuum and the soft excess.
The \mrk results show that the UV continuum and the soft excess vary simultaneously, and more strongly than the power-law, which are not expected in the reflection interpretation of the soft excess.
Interestingly, recent \swift reverberation mapping surveys of AGNs also suggest that the lamp-post reprocessing model, where variability of the central X-ray corona drives variability by the surrounding disk, is largely inconsistent with the observed correlations and lags (e.g. \citealt{Edel19}). 
While the trend of lags with wavelength broadly agrees with the reprocessing prediction \citep{Faus16}, towards longer lags the inconsistencies become more apparent.

The SED dimming of \mrk is, to some extent, reminiscent of the changing-look AGNs, where significant changes in accretion may occur.
In the case of changing-look NGC~3516 \citep{Mehd22a}, the intrinsic luminosities of the disk and the soft excess component, and the temperature of the disk, dropped together like in \mrk.
However, in the case of \mrk these SED changes are not as extreme, or long lived, as those in the changing-look AGNs.
Also, in changing-look NGC~3516 all components of the SED declined \citep{Mehd22a}, whereas in \mrk the X-ray power-law and the reflection component hardly changed.
Thus, the variability of \mrk is at an intermediate level between typical AGN variabilities and the more extreme ones that are associated to the changing-look AGNs. 
In a typical AGN, the shape of the optical/UV continuum does not alter significantly. 
However, in changing-look AGNs the continuum does change its shape, like in NGC~3516, and to a lesser extent in \mrk.

The observed variability behavior in \mrk is not necessarily indicative of a substantial and persistent change in the accretion rate, but rather it demonstrates that the warm corona (soft excess) is more closely associated to the disk (optical/UV) than the hot corona (power-law). This is likely because the warm corona is physically connected to the accretion disk, whereas the hot corona is more independent from the disk. The location and the formation of the hot corona is still debatable, and its association to the accretion disk may be less straightforward than that of the warm corona.

The results of our investigation highlight the importance of both optical/UV and X-rays for studying spectral hardening events in AGNs because without the UV spectral information one may erroneously associate the lowered X-ray flux to obscuration.
In the soft X-ray band alone, the decline in flux can mimic obscuration. 
However, by taking into account the entire SED it becomes apparent that the flux decline in \mrk is a multi-wavelength phenomenon, spanning from optical/UV to the soft X-rays.
The new \mrk results suggest that the soft X-ray excess component is inherently connected to the optical/UV continuum, being the high energy tail of the disk emission.
The warm Comptonization model for the soft excess is the only plausible model that can explain such optical/UV and X-ray variability characteristics.

\begin{acknowledgments}
Our NuSTAR, XMM-Newton, and Swift data were obtained and supported through the NuSTAR Guest Observer joint program, NASA grant 80NSSC22K0552. 
The NuSTAR mission is a project led by the California Institute of Technology, managed by the Jet Propulsion Laboratory, funded by the National Aeronautics and Space Administration. 
This work is also supported by NASA through a grant for HST program number 16905 from the Space Telescope Science Institute, which is operated by the Association of Universities for Research in Astronomy, Incorporated, under NASA contract NAS5-26555. 
We acknowledge the use of the UK Swift Science Data Centre at the University of Leicester.
We thank the anonymous referee for providing constructive comments and suggestions that improved the paper.
\end{acknowledgments}
\facilities{NuSTAR, XMM, Swift, HST (COS)}
\bibliographystyle{aasjournal}
\bibliography{references}{}
\end{document}